# DESORPTION OF HOT MOLECULES FROM PHOTON IRRADIATED INTERSTELLAR ICES

J. D. Thrower [1], D. J. Burke [2], M. P. Collings [1*], A. Dawes [3], P. J. Holtom [3], F. Jamme [4], P. Kendall [3], W. A. Brown [2], I. P. Clark [5], H. J. Fraser [6], M. R. S. McCoustra [1], N. J. Mason [3], and A. W. Parker [5]

(1) School of Engineering and Physical Sciences, Heriot-Watt University, Edinburgh, EH14 4AS, UK.

(2) Department of Chemistry, University College London, 20 Gordon Street, London, WC1H 0AJ, UK.

(3) Department of Physics and Astronomy, The Open University, Walton Hall, Milton Keynes, MK7 6AA, UK.

(4) School of Chemistry, University of Nottingham, University Park, Nottingham, NG7 2RD, UK. Permanent Address: SOLEIL Synchrotron, BP 48, L'Orme des Merisiers, F-91192 Gif sur Yvette Cédex, France.

(5) Central Laser Facility, Science and Technology Facilities Council, Rutherford Appleton Laboratory, Didcot, Oxon, OX11 0QX, UK.

(6) Department of Physics, Scottish Universities Physics Alliance (SUPA), University of Strathclyde, John Anderson Building, 107 Rottenrow East, Glasgow, G4 0NG, UK.

* Corresponding author: m.p.collings@hw.ac.uk

ABSTRACT

We present experimental measurements of photodesorption from ices of astrophysical relevance. Layers of benzene and water ice were irradiated with a laser tuned to an electronic transition in the benzene molecule. The translational energy of desorbed molecules was measured by time-of-flight (ToF) mass spectrometry. Three distinct photodesorption processes were identified – a direct adsorbate-mediated desorption producing benzene molecules with a translational temperature of around 1200 K, an indirect adsorbate-mediated desorption resulting in water molecules with a translational temperature of around 450 K, and a substrate-mediated desorption of both benzene and water producing molecules with translational temperatures of around 530 K and 450 K respectively. The translational temperature of each population of desorbed molecules is well above the temperature of the ice matrix. The implications for gas-phase chemistry in the interstellar medium are discussed.

*Subject Headings:* astrochemistry --- ultraviolet (ISM) --- ISM: molecules --- comets: general --- methods: laboratory

1. INTRODUCTION

Polycyclic aromatic hydrocarbons (PAHs) are thought to be ubiquitously present in the interstellar medium (ISM) (Hudgins & Allamandola 2005). Emission features attributable to PAHs are observed towards, for example, planetary nebulae (Lagadec et al. 2005), photodissociation regions (Kassis et al. 2006), diffuse interstellar clouds (Flagey et al. 2006) and extra-galactic sources (Smith et al. 2007). They have been extracted from chondritic meteorites (Hahn et al. 1988) and interplanetary dust particles (Clemmett et al. 1993). Where the temperature is cold enough to cause freeze-out of gas-phase molecules onto dust grains, PAHs are expected to be mixed with other molecules in icy mantles – absorption features attributed to PAHs are observed along lines of sight towards protostellar objects (Bregman & Temi 2001), and PAHs have recently been detected in cometary comas (Lisse et al. 2007). The spectroscopy and photochemistry of PAHs in the solid phase have been extensively studied (Gudipati & Allamandola 2004, Mattioda et al. 2005, and references therein). In this publication, we study photodesorption from astrophysically relevant ices after laser excitation of electronic transitions associated with the aromatic ring of an organic molecule. Here we have used benzene as a 'model' PAH, primarily for reasons of experimental convenience. However, the $\pi \rightarrow \pi^*$ transition that is excited in this work is common to all aromatic molecules, and lies at increasingly longer wavelengths as the molecular complexity increases (Silverstein et al. 1991).

Photodesorption is potentially an important process in a variety of astrophysical environments. It is a mechanism that is often invoked to account for the high gas-phase abundances of less volatile molecules, such as water, that are observed under conditions where complete freeze-out onto dust grains might otherwise be expected (Willacy & Williams 1993; Smith et al. 1993; Willacy & Langer 2000; Dominik et al. 2005). However, there are relatively few published experimental studies of photodesorption within an astrophysical context. Westley et al. (1995a, 1995b) measured desorption of water ice irradiated with Lyman α irradiation using microbalance techniques, and Öberg et al. (2007) have recently published a study of photodesorption of carbon monoxide under Lyman α radiation. For the most part, these previous studies have focussed on the measurements of the rate of photodesorption, whereas the experiments presented in this paper focus on the energetics of the process. Little attention has been given to experiments investigating UV photodesorption from more complex ices mimicking those that might be found in dust grain mantles within dense molecular clouds, or in cometary ices, that might address these questions. There are, however, a few published studies of IR laser promoted photodesorption from benzene, PAH and water ice films (Braun & Hess 1993; Mihesan et al. 2006; Fosca et al. 2003). In astrophysical environments photodesorption can result due to UV irradiation from stellar sources, and also from UV photons produced following cosmic ray excitation of gas-phase $H_2$ molecules. These mechanisms operate on different timescales, and are prevalent under different astrophysical conditions, and therefore are treated as separate processes. In the laboratory however, we can make no distinction between the source of photons, and therefore our results are relevant to both processes.

The results presented here are the first from a recently established collaborative project that has been designed to characterise the photon-driven physics and chemistry of PAHs and water ice in the ISM. Photon irradiation of astrophysical ices can have a variety of distinct consequences, including the formation of ions and radicals that result in chemical changes in the ice, the desorption of molecules as is demonstrated here, morphological or structural changes to the ice, heating of the ice, and re-emission of photons. The ultimate goal of the project is to fully quantify the distribution of photonic energy absorbed by such ices over a wide spectral range. In these initial experiments, we have not attempted to quantify the rate of the photodesorption process. However, quantitative analysis of this rate, in addition to simultaneous measurements of the rates of chemical and morphological change in the ice, is within the capabilities of the apparatus. Future experiments will therefore aim to measure the cross-sections for photodesorption and the various other consequences of photon irradiation. Such measurements will represent an important step forward in the understanding of the behaviour of photon irradiated ices in astrophysical environments.

## 2. EXPERIMENTAL PROCEDURES

The dedicated ultrahigh vacuum (UHV) chamber in the Central Laser Facility (CLF) at the Rutherford Appleton Laboratories (RAL) in which experiments were performed is described in detail elsewhere (Jamme et al. 2007, in preparation). It has a base pressure of approximately $1 \times 10^{-10}$ mbar, somewhat higher than the pressures of $10^{-12}$ mbar that typify the more dense regions of molecular clouds (van Dishoeck 1998). Analytical grade benzene and deionised water were further purified by repeated freeze-pump-thaw cycles, and introduced into the chamber via leak valves. Layers of benzene and water were deposited onto a sapphire crystal (1 mm × 10 mm diameter) at a temperature of roughly 80 K, by backfilling the chamber to a pressure of $4 \times 10^{-7}$ mbar for 500 s. The pressure was not corrected for ion gauge sensitivity. The thicknesses of the water and benzene layers are estimated using literature values of the densities of amorphous solid water and solid benzene (Kimmel et al. 2001, Craven et al. 1993), and assuming a sticking probability of unity, to be 16 nm and 2 nm respectively.

The films were irradiated with the frequency-doubled output of a nanosecond pulsed $Nd^{3+}$-YAG pumped dye laser. The laser beam was incident on the sapphire sample at an angle of 45° to the surface normal, and focussed to an area of approximately 0.5 mm². The laser was operated with a pulse frequency of 10 Hz, and pulse energies of either 1.8 mJ or 1.1 mJ, corresponding to irradiances of 360 or 220 mJ cm$^{-2}$ per pulse. The pulse-to-pulse variation in the irradiance was typically about ± 10 %. Molecules desorbing normal to the surface were detected by a pulse counting quadrupole mass spectrometer, after flying down a liquid nitrogen cooled line-of-sight tube (Jones & Turton 1997).

The benzene and water molecules were detected by their parent ions of mass 78 and 18 $m_u$, respectively. The time-of-flight (ToF) of the desorbed molecules was measured with a resolution of 2.56 µs over a range of 0 – 42 ms, by a multichannel scaler which was triggered after each laser pulse. Data was accumulated for 200 laser pulses, after which time the desorption rate had fallen below the limit of detection in all experiments. Data was typically averaged over 30 spots from different locations on the sample surface. The ToF profiles were fitted with a density weighted Maxwell-Boltzmann equation (Zimmerman & Ho 1995), to determine the translational temperature, $T_t$, of the desorbed molecules.

Experiments were conducted at three laser wavelengths. The 'on-resonance' wavelength of 250.0 nm was chosen to excite one of the several vibronic components of the $B_{2u} \leftarrow A_{1g}$ transition of pure benzene ice, a $\pi(1e_{1g}) \rightarrow \pi^*(1e_{2u})$ promotion of an electron in the aromatic ring of the benzene molecule. The widths and precise positions of the vibronic components of this transition are known to vary with the temperature of solid benzene, and also with the concentration of benzene in the water mixture (Mason et al. 2007, unpublished results). We have chosen 248.8 nm as a 'near-resonance' wavelength, as we believe it to be close to a local minimum in the absorption cross-section of benzene. However, solid benzene will certainly have a non-negligible absorption at this wavelength. The absorption cross-section of benzene should be negligible at the 'off-resonance' wavelength of 275.0 nm. Unfortunately, off-resonance experiments could not be conducted with the higher irradiance, due to the rapid accumulation of damage to the focussing mirrors under these conditions. The absorption cross-section of water ice is negligible over the entire wavelength range (Kobayashi 1983). However, sapphire has a small but significant absorbance over this range (Patel & Zaidi 1999). Based on these absorbances, it is estimated that at 250.0 nm, 0.1 % and 4 % of the photon flux is absorbed by the benzene layer and the sapphire substrate respectively. We therefore expect substrate-mediated processes due to heating of the sapphire to have an important influence on the results. The bandwidth of the laser was $6.25 \times 10^{-4}$ nm, very much narrower than the benzene absorption band.

With the experimental method adopted, desorption is measured until all of the ice under a laser spot has desorbed. Therefore, it might be assumed that the desorption peaks will have equal size, regardless of the rate of desorption, and thus also of the laser wavelength. In order to restrict the detection of molecules by the mass spectrometer to those which originate from the sample, the line-of-sight tube greatly limits the solid angle in which desorbed molecules can be detected. We estimate that only 0.01 % of molecules leaving the sample surface reach the ionization region of the mass spectrometer. In combination with the small spot size, this means that the mass spectrometer signal is rather low, and even under the most favourable conditions for desorption, the signal may not be far above the detection limit. As illustrated in figure 1, under such conditions, the integrated area of a desorption peak does not reflect the total number of molecules desorbed, but instead increases with increasing rate of desorption. Detection is therefore biased towards rapid processes.

## 3. RESULTS AND DISCUSSION

Figure 2 shows the ToF profiles for photodesorbed benzene and water when deposited onto the sapphire sample in varying layer configurations when irradiated by 250.0 nm photons. We denote the layer configurations as follows – benzene adsorbed alone: S/B; water adsorbed alone: S/W; benzene adsorbed followed by water: S/B/W; water adsorbed followed by benzene: S/W/B. A single peak is observed for desorbed benzene at approximately 0.5 ms in each configuration where it is present. The size of this peak is substantially reduced for the S/B/W system in comparison to the S/B system. In contrast, the peak is slightly enhanced in the S/W/B system. A peak is observed at roughly 0.35 ms for desorbed water in each of the configurations where water is present. Water was observed to desorb in the absence of benzene. We believe that this desorption can be attributed to a substrate-mediated mechanism, as discussed below. Water desorption is enhanced in the presence of benzene.

Therefore, an indirect adsorbate-mediated process must also be operating, in which energy absorbed by benzene molecules is transferred to water molecules causing them to desorb. The enhancement is slight for the S/B/W system, but much greater for the S/W/B system.

A possible reason for the variation in the observed rates of desorption is the morphology of the deposited films. Figure 3 illustrates the relevant aspects of the anticipated morphology in cartoon form. For the S/B/W system, the desorption of benzene is inhibited by the presence of the thick overlayer of water ice, hence the desorption rate is lower than for the S/B system. Amorphous water ice deposited at 80 K is known to have an uneven surface, resulting in a high surface area (Collings et al. 2003). Water is known not to wet the surface of graphite (Lofgren et al. 2003; Bolina et al. 2005), therefore it is likely that the same will be true of benzene adsorbing on water ice. The relatively thin layer of benzene clustering in islands may result in a dispersed overlayer with high surface area. This model is supported by the results of recently published experiments using temperature programmed desorption, infrared spectroscopy and metastable impact electron spectroscopy (Bahr & Kempter 2007). Thus, the rate of benzene desorption in the S/W/B system can be expected to be greater than that of the S/B system due to the larger surface area of benzene available. The rate of water desorption from the S/B/W system is greater than from the S/W system because of the contribution of the indirect adsorbate-mediated process. However, the excited water molecules must still escape from the benzene/water interface through the thick water overlayer. In the S/W/B system, the dispersed benzene overlayer poses much less of a barrier, and the rate of water desorption is correspondingly higher.

Figure 4 displays the wavelength dependence of the photodesorption of benzene from the S/B system and water from the S/W/B system. At the off-resonance wavelength, benzene desorption is evident, which again indicates that a substrate-mediated desorption mechanism is operative. A single component Maxwell-Boltzmann fit gives a translational temperature of 530 K. As expected, the benzene desorption is much more intense at the on-resonance wavelength. The on-resonance peak is also slightly faster than the off-resonance peak. This desorption peak is therefore thought to arise through a combination of the direct adsorbate-mediated and substrate-mediated desorption mechanisms. While a single component routine provides an adequate fit to each benzene ToF profile, we have applied a two component fitting routine to the on-resonance and near-resonance peaks in order to derive a temperature for the direct adsorbate-mediated mechanism. The temperature of one component was fixed at 530 K, while the temperature of the second component and the intensities of both components were allowed to vary. The size of the 530 K component was found to be similar in each experiment, as is expected for a substrate-mediated process which will be relatively insensitive to the photon wavelength, since the sapphire absorption cross-section itself is relatively constant in this wavelength region. For the on-resonance experiment, the free component which we assign to the adsorbate-mediated desorption of benzene dominates the ToF profile, having a much higher translational temperature of 1200 K. The two component fit to the near-resonance data produces a small peak for adsorbate-mediated desorption at 1030 K. After averaging the fitted temperatures weighted against the size of the component over all of the benzene desorption experiments, we find the translational temperature of benzene desorbed by the direct adsorbate-mediated and substrate-mediated mechanisms to be $1200 \pm 200$ K and $530 \pm 100$ K, respectively. A single component fit to the on-resonance water desorption from the S/W/B system gives a translational temperature of 450 K. However, at the lower laser power, the rates of water desorption at near-resonance and off-resonance shown in figure 4B were too low to obtain a meaningful temperature fit. Although we have concluded from the results in figure 2B that two mechanisms for water desorption are operating, no significant differences in the translational temperature were apparent. Water desorption by both indirect adsorbate-mediated and substrate-mediated mechanisms occur at a similar temperature, which when averaged over all of the experiments performed was found to be $450 \pm 100$ K.

A much slower water desorption peak was also evident in some experiments (not shown). It was only observed for the S/B/W system, and while its presence, position and intensity did not show good spot-to-spot reproducibility, it was evident in each of the on-resonance and near-resonance experiments for this system. It showed similar intensity to the rapid desorption peaks discussed above, and was centred at about 10 ms. A Maxwell-Boltzmann fit gives an unphysical translational temperature of less than 1 K. We believe that this peak may be due to the desorption of water clusters, which disintegrate within the mass spectrometer to contribute to the water monomer signal at mass 18 amu. The disintegration of such clusters may add an additional time delay, which when contributing to the apparent ToF of detected water molecules, prevents us from obtaining a sensible Maxwell-Boltzmann fit to the peak. Further experiments are required to confirm the desorption of water clusters.

Bergeld & Chakarov (2006) found that water is photodesorbed from a graphite surface in a multi-photon process when irradiated with photons in the 275 ~ 670 nm range. However, they report that desorption is limited to a population of molecules of not more than 0.25 monolayers which are weakly bound at the water ice surface, such that the overall desorption yield is independent of thickness. Given the lower sensitivity of our experiment, it seems unlikely that we would be able to detect such a small yield. In contrast, Nishi et al. (1984) observed photodesorption of water by a 2-photon process from a 200 μm thick film of water ice irradiated by a 248 nm laser, with no evidence of any limitation of the desorption yield. The translational temperature of the desorbed molecules was measured to be 770 K. The photon flux in our experiments was within the lower range used by Nishi et al. (1984). However, the absence of evidence of a population of water molecules desorbed with a translational temperature near to 770 K in our experiments suggests that the absorption cross-section for the 2-photon process is too low to result in significant desorption under our experimental conditions, where the film thickness is some 4 orders of magnitude lower. There is strong evidence for significant absorption by the substrate under our experimental conditions. Pitting of the sapphire surface is evident, and a contrast is observed in the photodesorption traces from sapphire and platinum substrates (Jamme et al. 2007, in preparation). We are therefore confident that the 530 K desorption of benzene, and the desorption of water in the absence of benzene can be attributed to a surface-mediated mechanism. Substrate-mediated desorption was not observed in the studies by either Bergeld & Chakarov (2006) or Nishi et al. (1984). In the former case, the photon flux used was much lower than in our experiments, and was below the threshold for such substrate-mediated desorption (Ellegaard & Schou 1997). In the latter case, thermal desorption of water due to laser heating of the substrate is eliminated by the much greater thickness of the ice layers.

At the much lower photon fluxes prevalent within molecular clouds, grain and ice mantle processes induced by the simultaneous absorption of two or more photons are not relevant. It should be possible to determine the photon order of the desorption processes by comparison of the desorption rates at different laser energies. However, the simultaneous operation of more than one desorption mechanism makes this analysis more complex. The results that we have presented in this publication are consistent with desorption via a single-photon process, and as such, are of relevance to interstellar environments. Nevertheless, additional flux dependent measurements of desorption rates are essential.

## 4. ASTROPHYSICAL IMPLICATIONS

The experimental results demonstrate three distinct mechanisms for photodesorption – desorption resulting by direct excitation of the adsorbate molecule (direct adsorbate-mediated), by excitation of the adsorbate molecule followed by transfer of energy to a neighbouring molecule within the ice matrix (indirect adsorbate-mediated) and by absorption by the substrate (substrate-mediated). Each of these mechanisms is of relevance to astrophysical situations. The sapphire crystal upon which we deposit benzene and water ice is not an ideal representation of an interstellar dust grain, and was chosen as a substrate for its thermal properties. Absorption of UV radiation by the sapphire crystal makes analysis of the ToF traces more complex, since the isolation of the photodesorption processes within the ice layers is more difficult. However, the identification of a substrate-mediated photodesorption mechanism in our results demonstrates that absorption of UV photons by silicates and metal oxides in dust grains is of significance to photodesorption in astrophysical environments. Likewise, complex ices containing UV absorbing species will exhibit both of the adsorbate-mediated desorption channels.

The translational temperature of both the benzene and the water molecules was found to be much higher than the temperature of the ice matrix from which they were photodesorbed. This is the case for photodesorption by each of the three mechanisms. Therefore, it is likely that suprathermal desorption is a general effect for photodesorption, regardless of the molecule desorbed or the wavelength of the UV photons. The injection into the gas-phase of molecules with temperatures well above the ambient temperature can be expected to have significant effects on gas-phase chemistry. Such hot molecules may be able to surmount otherwise insurmountable reaction barriers, allowing a range of chemical reactions not otherwise possible. The elevated temperatures of these molecules should certainly have a strong influence upon the rates of reactions in which they take part. Therefore, these results may imply significant deviations from current chemical models of regions where the photodesorption of ices is a significant process. Given that there are some similarities in the chemical changes induced by irradiation of astrophysical ice analogues with protons and UV photons (Bernstein et al. 2003), we speculate that cosmic ray and electron bombardment may also lead to the desorption of hot molecules from astrophysical ices.

The results presented here were obtained for a limited range of conditions. Further study of photodesorption from the benzene-water system over an expanded range of conditions is planned for the near future. The variation of photodesorption rate with photon flux must be more thoroughly assessed to confirm that the photodesorption mechanisms have a single photon dependence. Variations of the precise wavelength of the on-resonance mechanism are expected for benzene-water mixtures at different compositions and temperature. Furthermore, ices grown at varied temperatures may cause changes in photodesorption rates due to morphological effects. Similarly, studies of varied ice layer thicknesses may help to provide a greater understanding of the photodesorption rate effects that we have attributed to morphology. However, these variables are expected to influence the kinetics of photodesorption, rather than the dynamics of the process. Therefore we believe that the main conclusions which can be drawn from this research – that the operation of three distinct mechanisms each yielding photodesorbed molecules with suprathermal translational energies – will be valid over wide ranging conditions, including those prevalent in the ISM.

The interstellar radiation field (ISRF) shows a local minimum close to the on-resonance wavelength of 250 nm in these experiments, but increases with increasing wavelength to a peak at around 1 μm (Mathis et al. 1983). The $\pi \rightarrow \pi^*$ transitions of larger PAH molecules occur in this region of higher photon flux. The ISRF is greatly attenuated within dense molecular clouds, although relatively less so at around 1 μm than at 250 nm (Mathis et al. 1983). The UV radiation field at the disk surface around young stars may be several orders of magnitude greater than the ISRF (Herbig & Goodrich 1986). The cosmic ray induced radiation field within cold dense clouds is dominated by Lyman and Werner band photons (Sternberg et al. 1987) at a somewhat shorter wavelength than that used in these experiments. However, benzene and PAH molecules have strong absorption bands in the 100 – 200 nm range.

Photodesorption of a molecule demands that at least the energy equivalent to the thermal desorption of the molecule is deposited in the desorption co-ordinate. In principle, if an ice can absorb a photon of equivalent or larger energy than this desorption energy, then provided that desorption is competitive with relaxation channels, desorption will occur. The desorption energies of amorphous solid water (Fraser et al. 2001) and solid benzene (Bahr & Kempter 2007) are approximately 46 kJ mol$^{-1}$ and 42 kJ mol$^{-1}$ respectively, which correspond to a photons of 2.6 μm and 2.8 μm. The desorption energy of PAHs will inevitably increase as the molecule increases in size, therefore this limiting wavelength will decrease, shifting through the near infrared region. The translational energy of the desorbed molecule is limited to a maximum of the difference between the energy of the photon and desorption energy for the molecule. The statistical distribution of energy between the possible outcomes of photon absorption may vary as different transitions are activated, therefore photodesorption rates may vary at different wavelengths. Although we have experimented with only a single transition of benzene, there is no reason to assume that these results are not more generally relevant to the family of PAH molecules over a broad spectral range. Nonetheless, further experiments are essential, and will provide interesting results.

## 5. CONCLUSIONS

We have studied the photon stimulated desorption of molecules from layered deposits of benzene and water ice. Three distinct desorption mechanisms were identified. Firstly, excitation of the $B_{2u} \leftarrow A_{1g}$ transition of benzene led to direct adsorbate-mediated photodesorption of the benzene with a translational temperature of around 1200 K. Secondly, excitation of this transition led to indirect adsorbate-mediated photodesorption of water with a translational temperature of around 450 K, after transfer of energy from benzene to water. Finally, absorption by the sapphire substrate led to photodesorption of both benzene and water via a substrate-mediated process yielding molecules with translational temperatures of around 530 K and 450 K respectively, after transfer of energy from the substrate to the ice matrix. For each mechanism, the translational temperature of the desorbed molecules was much greater than the surface temperature. We therefore predict that the photon stimulated desorption of suprathermal molecules will have important consequences for gas-phase chemistry in astrophysical environments.

Future experiments to characterise the photodynamics of benzene and water ice in much greater detail are planned. The photon flux dependence of desorption will be rigorously quantified, and the influence of temperature, composition of mixtures, thickness, and morphology of the ice will be studied. This last issue is particularly interesting, given that studies of the electron-stimulated dissociation of amorphous water ice have demonstrated that ice morphology is of critical importance (Grieves & Orlando 2005). New substrates will be prepared in an effort to minimise the substrate-mediated desorption mechanism. The experimental program will also be expanded to study PAHs with $\pi \rightarrow \pi^*$ transitions at longer UV wavelengths, which are thought to be of direct relevance to interstellar ices (Hudgins & Allamandola 2006).

We gratefully acknowledge the Science and Technology Facilities Council for its financial support in developing the UHV facilities at the CLF. JDT and MPC acknowledge the support of the UK Engineering and Physical Sciences Research Council (EPSRC). DJB acknowledges the support of the Leverhulme Trust. MPC and FJ acknowledge the financial assistance of the University of Nottingham. AD acknowledges the support of the UK Particle Physics and Astronomy Research Council (PPARC) for a Research Fellowship. PJH and PK acknowledge the support of the ESF Program EIPAM. We thank Dr Heike Arnolds and Dr Serena Viti for helpful discussions. We thank the referee selected by the Astrophysical Journal for their insightful and constructive comments.

FIGURE CAPTIONS

Fig 1. – An illustration of why for a weak desorption signal, the integrated area of the desorption peak detected by the mass spectrometer may reflect the rate of desorption rather than the total number of molecules desorbed. The thick solid curve and the dashed curve represent the desorption profiles for rapid and slow desorption processes respectively. The total desorption yield, as given by the area under the curve, is equal for the two profiles. The desorption yield above the mass spectrometer detection limit is much greater for the rapid desorption (shaded region) than for the slower desorption (cross hatched region).

Fig 2. – *(Colour online)* ToF profiles of (A) benzene and (B) water desorption for varying layer configurations, laser energy = 1.8 mJ, $\lambda$ = 250.0 nm. Thin grey lines: raw data; thick lines: single component Maxwell-Boltzmann fits. Profiles have been offset for clarity.

Fig 3. – *(Colour online)* Cartoon summarising the expected morphology of the layers of ice in each of the deposition configurations studied (not to scale).

Fig 4. – *(Colour online)* (A) ToF profiles of benzene desorption from the S/B system at varying photon wavelength; laser power = 1.1 mJ; thin grey lines: raw data; thick lines: Maxwell-Boltzmann fits, one component fixed at $T_t$ = 530 K (blue), one free component (red) and the sum of the two components (black). (B) ToF profiles of water desorption from the S/W/B system at varying photon wavelength; laser power = 1.1 mJ; thin grey lines: raw data; thick lines: single component Maxwell-Boltzmann fits. Profiles have been offset for clarity.

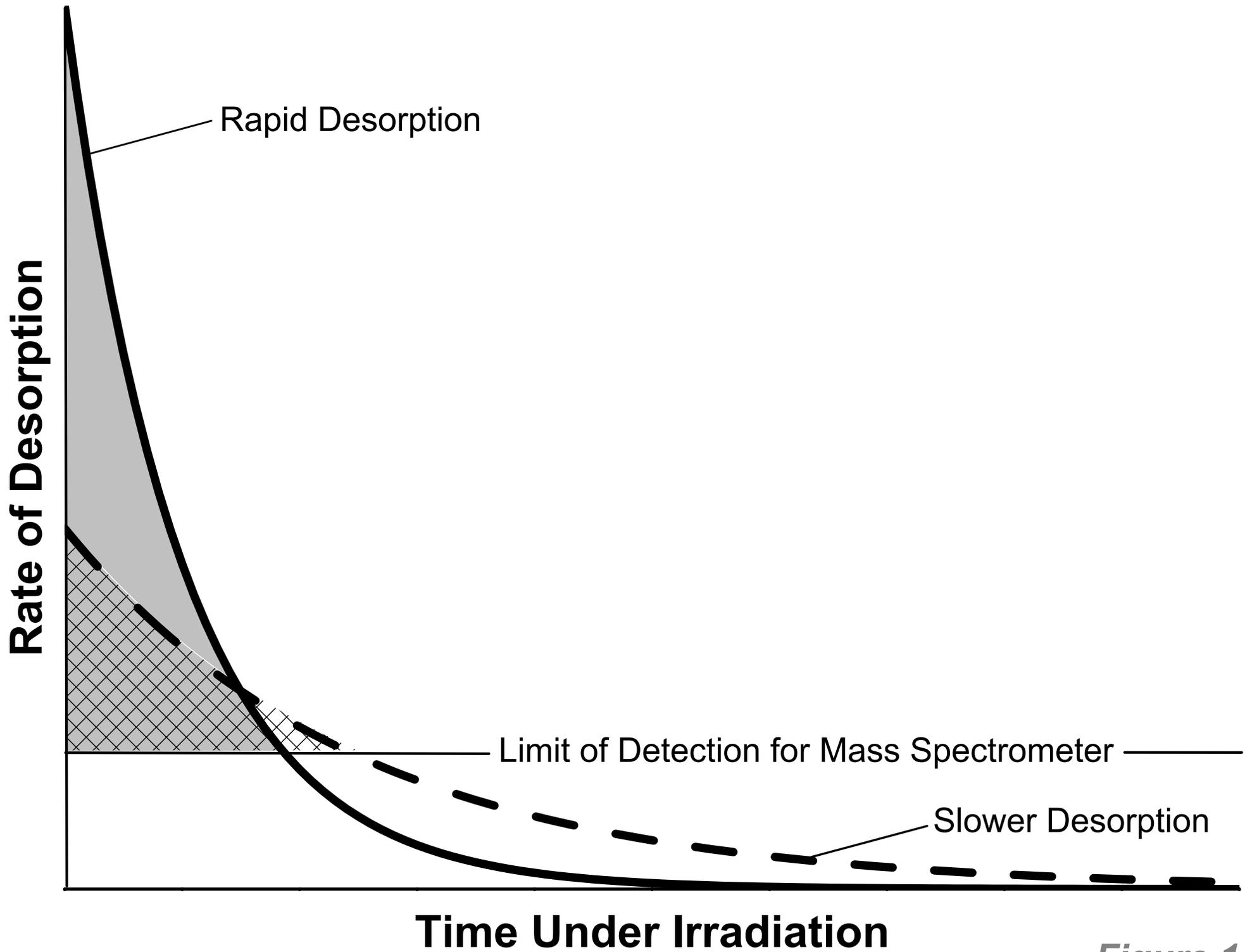

Figure 1

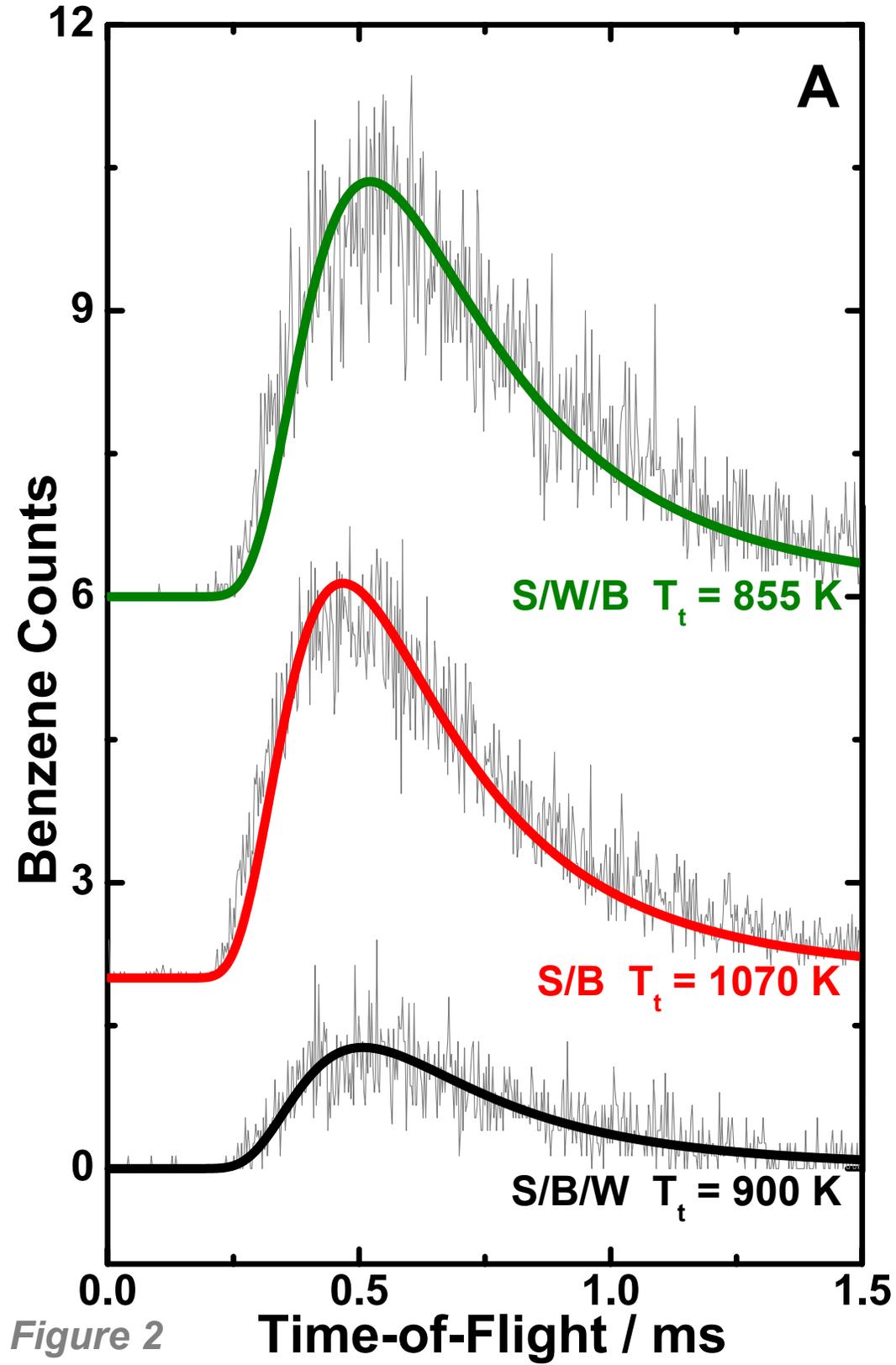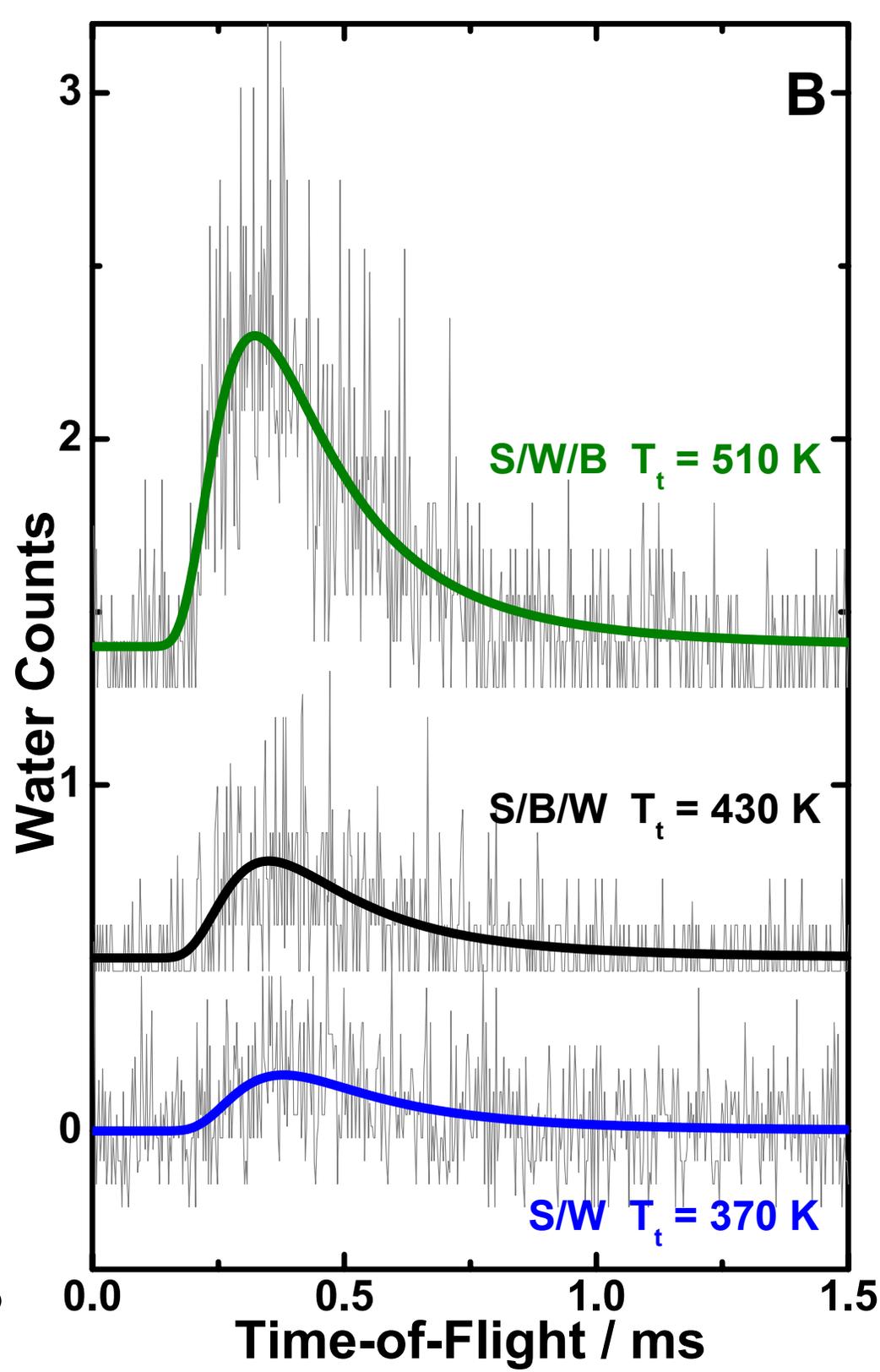

*Figure 2*

B
W
S
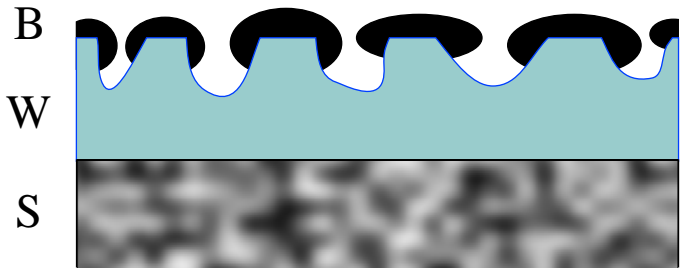
S/W/B: Water wets the surface of sapphire. The thinner layer of benzene does not wet the surface of water, and forms islands on its rough surface.

W
B
S
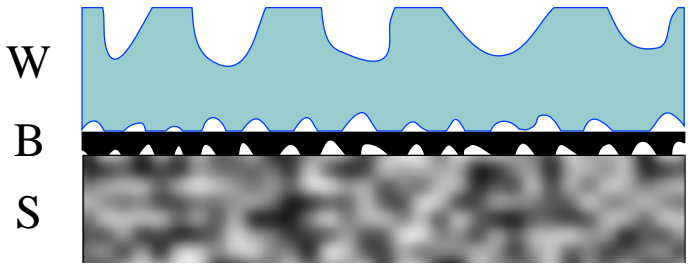
S/B/W: The thin layer of benzene does not wet the surface of the sapphire. The thicker overlayer of water does not wet the surface of the thin benzene layer.

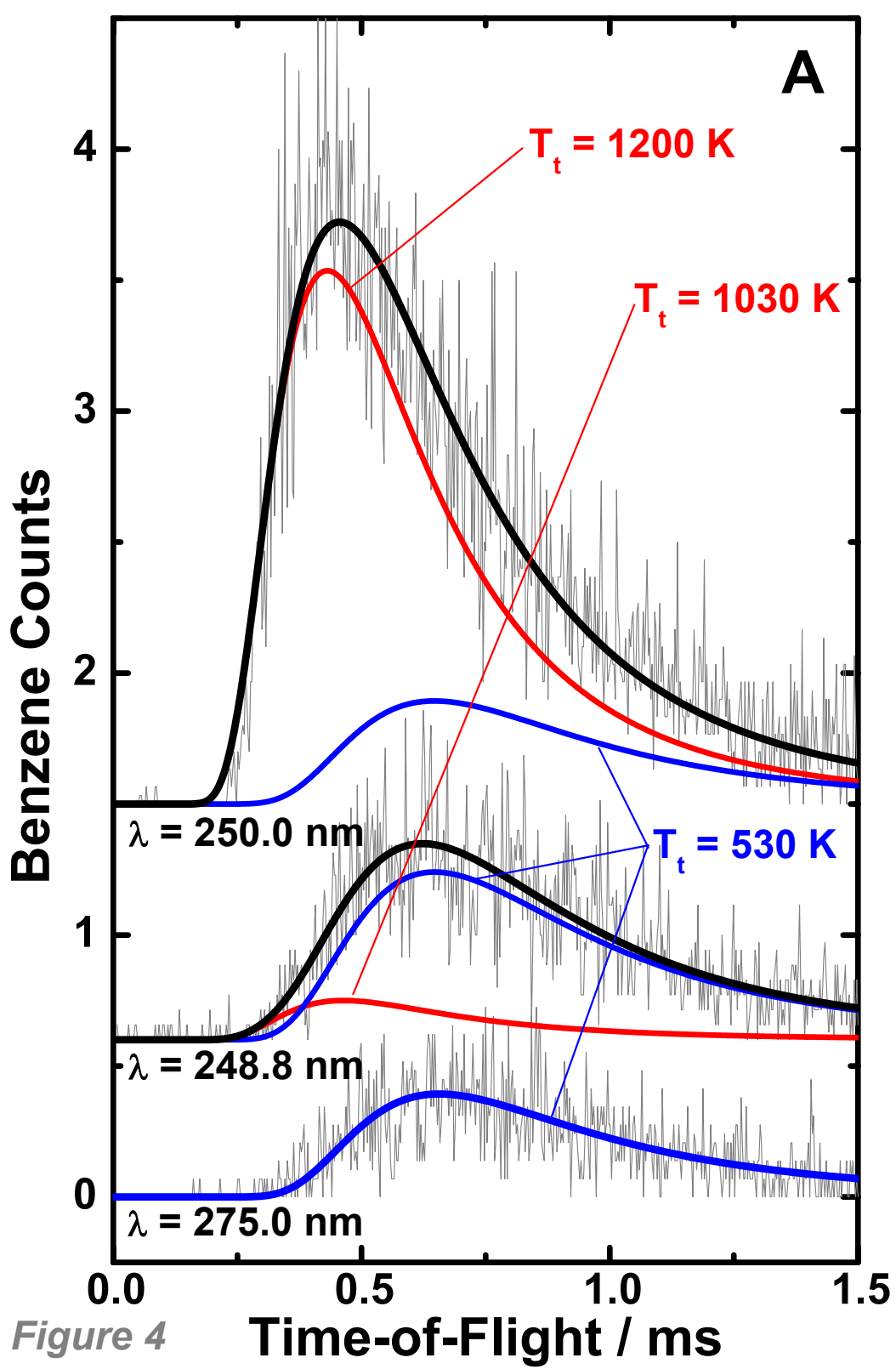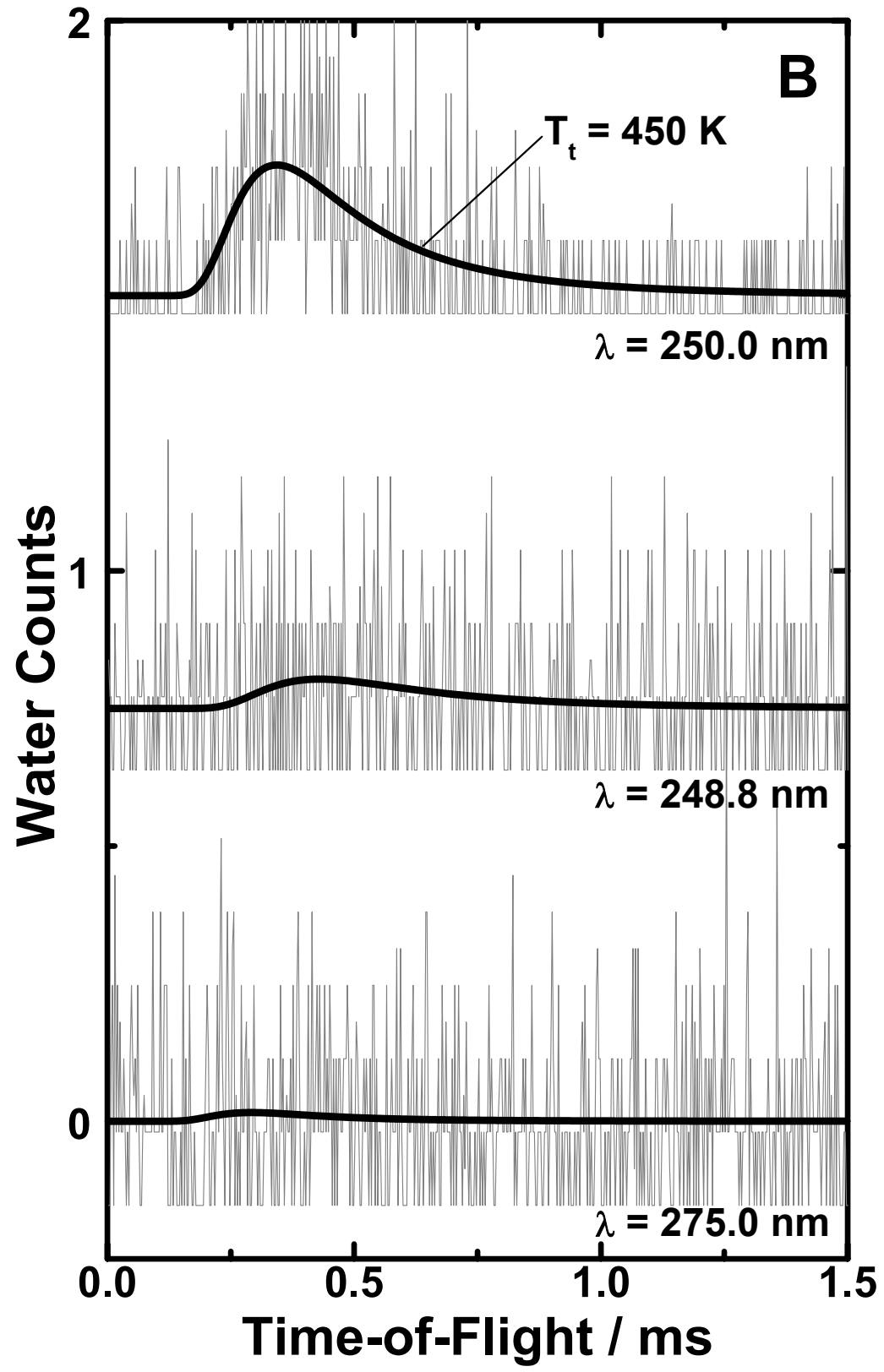

*Figure 4*